\documentclass[conference]{IEEEtran}

\usepackage{graphicx}   
\usepackage{booktabs}   
\usepackage{textcomp}   

\usepackage{tikz}
\usepackage{xcolor}
\usepackage{amsmath}

\usetikzlibrary{
  positioning,               
  arrows.meta,               
  calc,                      
  fit,                       
  shapes,                    
  shapes.geometric,          
  shapes.misc,               
  patterns,                  
  decorations.pathreplacing, 
  decorations.markings,      
  backgrounds                
}

\definecolor{accenthw}{RGB}{27,71,112}   
\definecolor{accentsw}{RGB}{176,96,20}   

\tikzset{
  >={Stealth[length=1.9mm,width=1.5mm]},
  block/.style={
    draw, thick, rounded corners=1.5pt, fill=black!3,
    align=center, font=\footnotesize, inner sep=3pt,
    minimum height=7mm, minimum width=10mm},
  bigbox/.style={
    draw, thick, rounded corners=3pt, fill=black!1, inner sep=6pt},
  hwreg/.style={
    draw=accenthw, thick, rounded corners=1pt, fill=accenthw!16,
    align=center, font=\footnotesize, inner sep=2pt,
    minimum height=6mm, minimum width=9mm,
    postaction={pattern=north east lines, pattern color=accenthw!45}},
  swreg/.style={
    draw=accentsw, thick, rounded corners=1pt, fill=accentsw!16,
    align=center, font=\footnotesize, inner sep=2pt,
    minimum height=6mm, minimum width=9mm},
  greg/.style={
    draw=black!55, rounded corners=1pt, fill=black!4,
    align=center, font=\footnotesize, inner sep=2pt,
    minimum height=6mm, minimum width=9mm},
  periph/.style={
    draw=black!60, thick, rounded corners=1.5pt, fill=black!7,
    align=center, font=\footnotesize, inner sep=3pt,
    minimum height=7mm, minimum width=10mm},
  fpgabound/.style={
    draw=black!70, dashed, line width=0.9pt, rounded corners=4pt,
    inner sep=8pt},
  arrowstyle/.style={->, thick, draw=black!75},
  sig/.style={arrowstyle},
  bus/.style={->, line width=1.3pt, draw=black!78},
  fwd/.style={->, thick, draw=accenthw, >=Stealth},
  stall/.style={->, thick, dashed, draw=accentsw, >=Stealth},
  note/.style={font=\scriptsize\itshape, align=center, text=black!62},
  lbl/.style={font=\scriptsize, align=center},
  taglbl/.style={font=\scriptsize\bfseries, align=center},
  dimline/.style={|<->|, draw=black!72, >=Stealth, thin},
  legend/.style={
    draw=black!40, rounded corners=2pt, fill=white,
    inner sep=4pt, font=\scriptsize},
  latch/.style={
    draw=black!65, fill=black!14, minimum width=3.2mm,
    minimum height=13mm, inner sep=0pt},
}

\newlength{\figcol}
\newlength{\figfull}

\begin{document}

\title{Zero-Instruction Sensor Reads: Register-Mapped Peripherals
and Hardware PWM on a Five-Stage Soft Processor}

\author{\IEEEauthorblockN{Nathanael Ren}
\IEEEauthorblockA{Duke University\\
Durham, NC, USA}}

\maketitle

\begin{abstract}
We present a case study in application-driven specialization of a
five-stage soft processor, evaluated on the inner control loop of a
reaction-wheel self-balancing bicycle. Starting from a custom 32-bit
RISC core in the MIPS tradition---five pipeline stages, full operand
forwarding with a single-cycle load-use interlock, and a 33.33~MHz clock
on an Artix-7 FPGA---we specialize the design in two ways. First, two
frequently accessed peripheral inputs are mapped directly into
architectural register state, written every cycle by hardware and owned
exclusively through the register file's write-port structure rather than
by arbitration. Second, four periodic pulse-width-modulation (PWM)
channels are offloaded to hardware and driven continuously from four
exported registers, removing periodic actuation from software entirely.
Because peripheral values are addressable as ordinary register operands,
all ten sensor reads in the control loop cost no dedicated instruction
and no dedicated cycle, folding into arithmetic that executes anyway;
the memory-mapped equivalent requires an explicit load per snapshot and
costs five extra instructions and cycles. The actuation path likewise
removes waveform maintenance from software entirely. We
report two configurations, because the extensions and the single-cycle
array multiplier they were deployed alongside are not present together
in a single archived build: an \emph{archived} configuration, whose
worst-case loop is 91~cycles ($2.73~\mu$s), and the \emph{integrated}
configuration matching the deployed system, at 43~cycles
($1.29~\mu$s). Against a 20~ms actuation frame these are margins of
roughly $7{,}300\times$ and $15{,}000\times$. The deadline is met by so
wide a margin in either case that the specialization was not necessary
for real-time compliance; its value lies in instruction count and
software simplicity, not in determinism, which an on-chip single-cycle
I/O region already provides. The zero-instruction sensor read is
independent of that choice: the multiplier cannot affect whether a
peripheral read needs an instruction of its own. The
contribution is a carefully measured account of what this design point
does, and does not, buy.
\end{abstract}

\section{Introduction}
\label{sec:intro}

Soft processors---general-purpose cores synthesized onto
reconfigurable logic---are attractive for embedded control because they
place the programmable core and its peripherals on the same die, under
the designer's full control. Mature ecosystems such as MicroBlaze,
Nios~II, and the many RISC-V soft cores expose peripherals through
memory-mapped I/O (MMIO): the processor reads a sensor or writes an
actuator by issuing a load or store to a reserved address, and a bus
fabric routes the access to the peripheral. MMIO is general, scales to
thousands of devices, and is the default for good reason. But it is not
free. Every peripheral read is a distinct load instruction, and on an
in-order pipeline with a load-use interlock that load can also cost a
stall cycle before its result is usable.

\begin{figure}[t]
  \centering
  \resizebox{\columnwidth}{!}{
\begin{tikzpicture}[scale=0.94, transform shape,
                    node distance=4mm and 8mm, font=\footnotesize]

  \node[hwreg] (r1) {\$1};
  \node[hwreg, right=6mm of r1] (r2) {\$2};

  \node[swreg, below=11mm of r1] (r26) {\$26};
  \node[swreg, right=2.5mm of r26] (r27) {\$27};
  \node[swreg, right=2.5mm of r27] (r28) {\$28};
  \node[swreg, right=2.5mm of r28] (r29) {\$29};

  \node[bigbox, fill=none, fit=(r1)(r2)(r26)(r29), inner sep=6pt] (rf) {};

  \node[bigbox, fill=none, fit=(rf), inner sep=10pt] (cpu) {};

  \node[lbl, rotate=90, anchor=south] at (cpu.west) {Processor};
  \node[lbl, rotate=90, anchor=south] at (rf.west) {Register File};

  \node[periph, above=12mm of r1] (joy) {Joystick\\(4-bit)};
  \node[periph, above=12mm of r2] (sen) {Sensor\\(16-bit)};

  \draw[sig] (joy) -- (joy |- cpu.north) -- (r1);
  \draw[sig] (sen) -- (sen |- cpu.north) -- (r2);

  \node[note, right=4mm of sen, anchor=west, text width=24mm]
        {written by hardware, every cycle};

  \node[block, below=12mm of r26] (p0) {PWM};
  \node[block, below=12mm of r27] (p1) {PWM};
  \node[block, below=12mm of r28] (p2) {PWM};
  \node[block, below=12mm of r29] (p3) {PWM};

  \draw[sig] (r26) -- (r26 |- cpu.south) -- (p0);
  \draw[sig] (r27) -- (r27 |- cpu.south) -- (p1);
  \draw[sig] (r28) -- (r28 |- cpu.south) -- (p2);
  \draw[sig] (r29) -- (r29 |- cpu.south) -- (p3);

  \node[note, left=5mm of p0, anchor=east, text width=22mm]
        {written by software, one instruction};

  \node[periph, minimum width=44mm, below=10mm of p1.south, anchor=north,
        xshift=6mm] (act) {Servo / Motor};

  \draw[sig] (p0) -- (p0 |- act.north);
  \draw[sig] (p1) -- (p1 |- act.north);
  \draw[sig] (p2) -- (p2 |- act.north);
  \draw[sig] (p3) -- (p3 |- act.north);

  \node[legend, anchor=north west, text width=40mm] at ($(act.south west)+(-24mm,-4mm)$) {
    \tikz\node[hwreg, minimum width=6mm, minimum height=3.5mm]{}; \, hardware-owned (\$1,\,\$2)\\[2pt]
    \tikz\node[swreg, minimum width=6mm, minimum height=3.5mm]{}; \, software-written (\$26--\$29)};

\end{tikzpicture}}
  \caption{The software's view of the machine. Two peripheral
  inputs are mapped into architectural registers that hardware writes
  every cycle ($\$1$, $\$2$); four registers written once by software
  ($\$26$--$\$29$) are exported continuously to the hardware PWM units
  that drive the actuators. Reading a sensor is reading a register;
  commanding an actuator is writing a register.}
  \label{fig:software_view}
\end{figure}
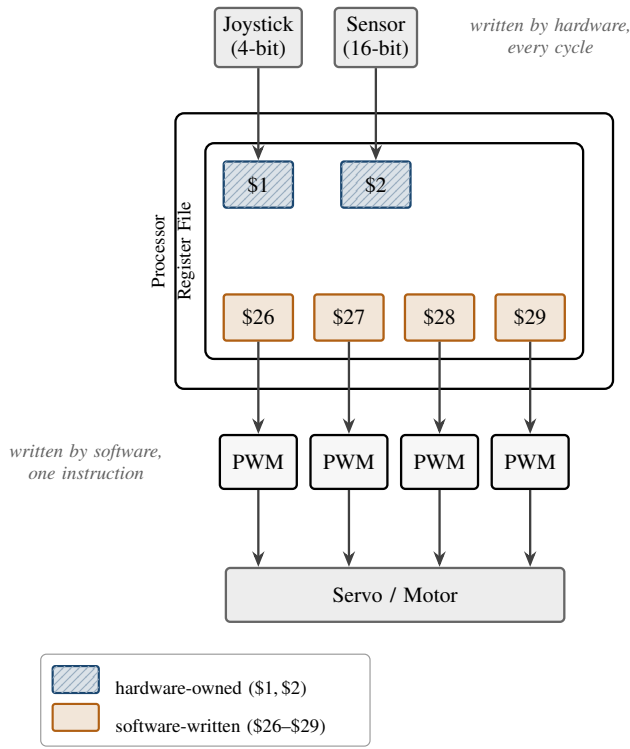

When the same programmable fabric hosts both the core and the
peripherals, a designer is free to specialize the interface between
them. This paper is a case study in doing exactly that, and in measuring
the result. We take a custom five-stage 32-bit RISC core in the
MIPS tradition and specialize it for one application, a reaction-wheel
self-balancing bicycle, in two ways. First, we map two frequently
accessed peripheral inputs directly into architectural register state: two
general-purpose registers are converted into dedicated hardware-written
flip-flop banks that re-capture their peripheral inputs every cycle, so
that reading a sensor is reading a register. Second, we offload periodic
actuation to hardware: four pulse-width-modulation (PWM) generators run
continuously in logic, each driven by a register the software writes
once, so that commanding an actuator is writing a register.
Figure~\ref{fig:software_view} shows this software view of the machine.

The contribution of this paper is a careful, quantified account
of what this specialization buys on a real design point, and what it does
not.

The answer has two parts. On the benefit side, at the instruction level
the register mapping is \emph{free}. Because a sensor value is an
ordinary register, it can serve
as an operand field inside an arithmetic instruction that would execute
anyway; in the control loop studied here, all ten peripheral reads and
one of the actuator writes disappear into existing instructions and cost
zero dedicated cycles. A memory-mapped interface structurally cannot
reach zero: it requires a distinct load instruction to bring the value
into a register before it can be used. Measured against the correct
reference loop, the memory-mapped equivalent costs five extra
instructions and five extra cycles, about $11.6\%$ per iteration, even
when a competent programmer snapshots each input once and schedules
around the load-use interlock. The actuation offload contributes a
qualitative simplification that is harder to price but real: no deadline
counters in the loop, no software waveform maintenance, no peripheral
base-address register, and no address arithmetic in the loop.

On the cost side, and this is the finding we consider most important,
the specialization was not necessary. We report two configurations,
because the extensions and the single-cycle array multiplier alongside
which they were deployed are not present together in a single archived
build: the \emph{archived} configuration, whose worst-case loop is
91~cycles ($2.73~\mu$s), and the \emph{integrated} configuration
matching the deployed system, at 43~cycles ($1.29~\mu$s) at the core's
33.33~MHz clock. The actuation deadline is one 20~ms PWM frame, so the
loop consumes roughly $0.0137\%$ or $0.0065\%$ of its deadline: margins
of about $7{,}300\times$ and $15{,}000\times$. The zero-instruction
sensor read holds in both, since the multiplier cannot affect whether a
peripheral read needs an instruction of its own. At either margin, an ordinary memory-mapped
implementation on the same core would also have met the deadline with
room to spare. The value of the specialization is not real-time
compliance, and it is not determinism, since an on-chip single-cycle I/O
region is already deterministic. What it buys is the instruction-count
reduction and the software simplicity described above. We state this
counterargument plainly because it is the strongest one against the work.

The remainder of this paper is organized as follows.
Section~\ref{sec:related} surveys background and related work.
Section~\ref{sec:baseline} describes the baseline core: its ISA,
pipeline, forwarding and hazard behavior, and arithmetic units.
Section~\ref{sec:extensions} presents the two specializations---the
register-mapped input interface and the hardware PWM actuation
path---and the register-file structure that enforces hardware ownership
without arbitration. Section~\ref{sec:evaluation} evaluates the control
loop cycle by cycle, establishes the zero-cost peripheral-access result,
compares it against a memory-mapped counterfactual, and reports the
timing-headroom finding. Section~\ref{sec:platform} briefly describes the
physical platform. Section~\ref{sec:discussion} discusses limitations,
and Section~\ref{sec:conclusion} concludes with future work.

\section{Background and Related Work}
\label{sec:related}

We acknowledge prior art specifically, at the level of \emph{access mechanism}.

\subsection{Register-Mapped Peripherals in Microcontrollers}
\label{sec:related-sfr}

Register-mapping peripherals is standard microcontroller practice since the
1980s. The Intel MCS-51 (8051) exposes ports, timers, and serial
peripherals as special-function registers in a dedicated address
space~\cite{intel8051}; the Atmel AVR uses dedicated \texttt{in}/\texttt{out}
instructions to a separate I/O register
space~\cite{AVR_InstructionSet_Manual}; Microchip PIC follows
suit~\cite{pic_midrange}. The mechanism is an \emph{I/O space fetched by an
explicit instruction}, not a GPR hardware writes.

\subsection{Peripheral and Coprocessor Interfaces in Soft Processors}
\label{sec:related-soft}

Closer are FPGA soft-processor interfaces. Xilinx MicroBlaze provides Fast
Simplex Link (FSL), later AXI4-Stream, channels reached by
\texttt{get}/\texttt{put} instructions~\cite{AMD_UG984_MicroBlaze}: a
\emph{dedicated instruction reading a stream FIFO}. Intel/Altera Nios~II
custom instructions wire \texttt{dataa}/\texttt{datab} operands and a
\texttt{result} port to the register
file~\cite{Altera_NiosII_CustomInstruction}---the nearest relative of our
design, but an instruction feeding a functional unit, not a register
reading back a live value. Tensilica Xtensa TIE adds ports, queues, and
state registers as architecturally-visible I/O~\cite{xtensa_tie}; the ARM
coprocessor interface (\texttt{MRC}/\texttt{MCR}, CP15) reaches non-core
state by dedicated instructions~\cite{arm_arm}. Each uses a dedicated
instruction or side channel; none an ordinary GPR operand.

\subsection{Architecturally-Visible State Beyond the GPRs}
\label{sec:related-csr}

RISC-V control and status registers (CSRs) form a separate register space
accessed by \texttt{csrr}/\texttt{csrw}
instructions~\cite{RISCV_Unprivileged_ISA, RISCV_Privileged_ISA};
hardware-updated CSRs (\texttt{cycle}/\texttt{time}) are the closest
analogue---hardware writes visible state---but in the 12-bit CSR space, not
the GPR file. Rocket Chip's Rocket Custom Coprocessor (RoCC) interface attaches
accelerators via custom instructions~\cite{Asanovic2016RocketChip}, and
transport-triggered architectures~\cite{corporaal1997tta} make computation
\emph{be} data movement to functional-unit ports. Closest in name, Biswas
et al.\ add \emph{architecturally visible storage}: compiler-managed storage
private to an instruction-set-extension accelerator, reached through the
extension instruction~\cite{Biswas2007ArchVisibleStorage}---not a GPR a
sensor overwrites and any instruction reads as an operand.

\subsection{Positioning}
\label{sec:related-positioning}

We claim no new mechanism. Our narrow difference: a GPR that hardware
rewrites every cycle, read as an ordinary operand with \emph{no access
instruction at all}---so access cost can reach the zero-instruction floor
the mechanisms above cannot (Section~\ref{sec:evaluation}). A focused search
surfaced no published description of it, likely reflecting narrowness, not
originality: 32 GPRs cap it at a few peripherals. The contribution is
quantifying this on a real-time workload, negative finding and all.

\section{Baseline Core}
\label{sec:baseline}

The specializations of Section~\ref{sec:extensions} are built on a
custom 32-bit RISC core designed from scratch. This section documents
the baseline so that the cost and effect of the specializations can be
attributed precisely. The core follows the MIPS tradition~\cite{patterson2013coa} in its
instruction encoding and register model but is an independent
implementation with its own conventions; where those conventions differ
from MIPS, we say so, because the differences are what make the
specializations inexpensive.

\subsection{Instruction Set}
\label{sec:baseline-isa}

The ISA comprises eleven opcodes plus eight R-type ALU operations.
Instructions are 32 bits with a 5-bit opcode field. R-type instructions
carry three 5-bit register fields and a 5-bit ALU-operation selector;
the eight ALU operations are add, subtract, bitwise AND, bitwise OR,
logical left shift, arithmetic right shift, multiply, and divide.
Immediate-type instructions carry a 17-bit immediate that is
sign-extended to 32 bits, and jump-type instructions carry a 27-bit
target. The opcode set covers register-register and register-immediate
arithmetic, loads and stores, conditional branches (branch-not-equal,
branch-less-than), unconditional and register-indirect jumps, jump-and-link,
and two instructions for exception state. Register fields are 5 bits,
naming 32 general-purpose registers.

There is no C compiler and no standard application binary interface for
this core. All software is hand-written assembly, and the register
conventions are the author's own rather than the MIPS ABI. This point
matters for the rest of the paper. Because no compiler and no calling
convention depend on the general-purpose registers, reserving specific
registers ($\$1$, $\$2$, and $\$26$--$\$29$) for dedicated hardware use
costs nothing: there is no ABI contract to violate and no allocator to
constrain. A reviewer who assumes MIPS conventions might expect these
reservations to be expensive; under this design's hand-written-assembly
model they are free.

\subsection{Pipeline}
\label{sec:baseline-pipeline}

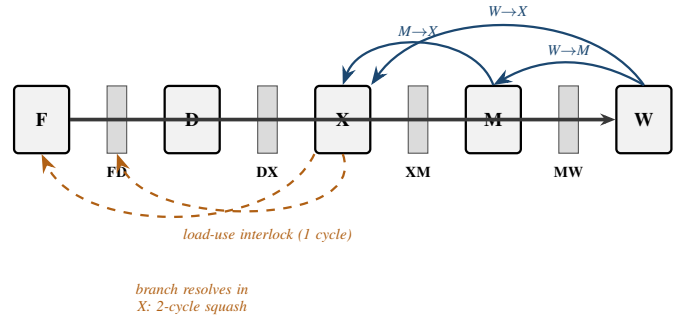
\begin{figure}[t]
  \centering
  \resizebox{\columnwidth}{!}{
\begin{tikzpicture}[scale=0.76, transform shape,
                    font=\footnotesize, node distance=0pt]

  \def\stagew{9mm}   
  \def\stageh{11mm}  
  \def\gap{6mm}      

  \tikzset{
    stage/.style={block, minimum width=\stagew, minimum height=\stageh,
                  fill=black!5, font=\small\bfseries},
    plbar/.style={latch, minimum height=\stageh}}

  \node[stage] (F) {F};
  \node[plbar, right=\gap of F] (FD) {};
  \node[stage, right=\gap of FD] (D) {D};
  \node[plbar, right=\gap of D] (DX) {};
  \node[stage, right=\gap of DX] (X) {X};
  \node[plbar, right=\gap of X] (XM) {};
  \node[stage, right=\gap of XM] (M) {M};
  \node[plbar, right=\gap of M] (MW) {};
  \node[stage, right=\gap of MW] (W) {W};

  \node[taglbl, below=1mm of FD] {FD};
  \node[taglbl, below=1mm of DX] {DX};
  \node[taglbl, below=1mm of XM] {XM};
  \node[taglbl, below=1mm of MW] {MW};

  \draw[bus] (F) -- (W);

  \draw[fwd] (M.north) to[out=110, in=70] node[note, above, pos=0.5, text=accenthw]{M$\to$X} (X.north);
  \draw[fwd] (W.north) to[out=125, in=55, looseness=0.9]
        node[note, above, pos=0.5, text=accenthw, yshift=1pt]{W$\to$X} (X.north east);
  \draw[fwd] (W.north) to[out=150, in=30]
        node[note, above, pos=0.5, text=accenthw]{W$\to$M} (M.north);

  \draw[stall] (X.south) to[out=-70, in=-70, looseness=0.9] (FD.south);
  \node[note, text=accentsw, text width=34mm, anchor=north]
        at ($(DX |- X.south) + (0,-11mm)$)
        {load-use interlock (1 cycle)};

  \draw[stall] (X.south west) to[out=-115, in=-65, looseness=0.85] (F.south);
  \node[note, text=accentsw, text width=34mm, anchor=north]
        at ($(D |- X.south) + (0,-20mm)$)
        {branch resolves in X: 2-cycle squash};

\end{tikzpicture}}
  \caption{Five-stage pipeline (F/D/X/M/W) with four latches. Forwarding
  covers M$\rightarrow$X and W$\rightarrow$X on both ALU operands and
  W$\rightarrow$M for store data. The load-use hazard is the only RAW
  case not covered by forwarding and costs exactly one interlock cycle.
  Branches resolve in execute under a predict-not-taken policy, costing
  a two-cycle squash when taken and nothing when not taken.}
  \label{fig:pipeline}
\end{figure}

The core is a five-stage in-order pipeline---fetch (F), decode (D),
execute (X), memory (M), and writeback (W)---separated by four
inter-stage latches (FD, DX, XM, MW). Figure~\ref{fig:pipeline} shows the
structure together with the exact forwarding and hazard behavior
described below. In steady state the pipeline sustains one instruction
per cycle.

\subsection{Forwarding and Hazards}
\label{sec:baseline-hazards}

Operand forwarding covers M$\rightarrow$X and W$\rightarrow$X on both ALU
operands, plus W$\rightarrow$M for store data. This network resolves
every read-after-write hazard except the load-use case at zero cost: an
arithmetic producer's result is available in the execute stage's input
whether the consumer is one or two slots behind it.

The single exception is a load feeding an immediately dependent
instruction. A loaded word only reaches the writeback stage; while the
value is in transit, the memory-stage register holds the load's
\emph{address}, not its data. A dependent instruction exactly one slot
behind the load therefore cannot be satisfied by the M$\rightarrow$X
path, because the value it needs is not yet where forwarding could take
it from. The hardware detects this case and interlocks for exactly one
cycle: it freezes the program counter and the FD latch and injects a NOP
into the decode-to-execute latch. After the single bubble, the loaded
value has advanced to writeback and the W$\rightarrow$X path delivers it.
The load-use penalty is thus exactly one cycle, and it is the only stall
the forwarding network cannot eliminate.

\subsection{Branches}
\label{sec:baseline-branches}

Branches resolve in the execute stage under a static predict-not-taken
policy. Fetch proceeds sequentially; when a branch is found to be taken,
the two younger instructions already fetched behind it are squashed to
NOPs. A taken branch therefore costs two squash cycles, and a not-taken
branch costs nothing. There is no dynamic branch prediction, no branch
target buffer, and no delay slot. Resolving branches late, in execute
rather than decode, is a deliberate choice: it lets branch and
jump-register operands reuse the same forwarding network as arithmetic
instructions rather than requiring a separate bypass into decode.

\subsection{Arithmetic Units}
\label{sec:baseline-alu}

The adder is a two-level 32-bit carry-lookahead adder built from four
8-bit carry-lookahead blocks, with a second level of lookahead computing
the inter-block carries from per-block generate and propagate signals.
The shifter is a five-stage logarithmic barrel shifter: logical
left-shift and arithmetic right-shift are realized by five fixed-shift
stages (16, 8, 4, 2, and 1 positions), each gated by one bit of the
shift amount.

Multiplication is performed by a single-cycle combinational array
multiplier in the form of a Wallace tree. The array is Python-generated and uses Baugh--Wooley sign
correction; it comprises 32 rows of full adders (993 full-adder instances
in all), with carries propagating within each row and each row consuming
the previous row's partial sum. Because the multiplier is combinational
and single-cycle, a multiply does not stall the pipeline. Division is
performed by a separate non-restoring divider that is multi-cycle
(approximately 32 cycles) and blocking: a divide freezes the pipeline
latches until it completes.

\subsection{Exceptions}
\label{sec:baseline-exceptions}

The core supports arithmetic overflow exceptions only. Overflow on an
add, add-immediate, subtract, multiply, or divide produces an exception
code (1 through 5, respectively) that is written to register~$\$30$,
which serves as the exception-status register. The exception-state
instructions ($\mathtt{setx}$ and $\mathtt{bex}$) operate through
$\$30$: $\mathtt{setx}$ writes a value into it, and $\mathtt{bex}$
branches when it is nonzero. In the steady-state control loop this
mechanism is never triggered.

\subsection{Clocking and Target Device}
\label{sec:baseline-clock}

The core runs at 33.33~MHz, a 30~ns clock period, derived from the
100~MHz board oscillator by a divide-by-three PLL. The design targets a
Xilinx Artix-7 XC7A100T on a Nexys~A7-100T board and is built with
Vivado. The 30~ns period is not incidental: as Section~\ref{sec:pwm}
shows, the PWM frame arithmetic only produces the correct 20~ms servo
frame at this clock. The design constraint file specifies only the
100~MHz input clock; the 33.33~MHz core domain is derived by the
clocking IP.

\section{Specializations}
\label{sec:extensions}

We specialize the baseline core in two ways: two general-purpose
registers are turned into hardware-written inputs
(Section~\ref{sec:regmap}), and periodic actuation is offloaded to four
hardware PWM generators driven by exported registers
(Section~\ref{sec:pwm}). Both specializations rely on one detail of the
design's clocking scheme, which we state first because it is the
mechanism that makes a hardware write visible to a software read within
the same cycle.

\subsection{Clocking Scheme}
\label{sec:clocking}

The datapath state---the program counter and the four pipeline
latches---updates on the negative edge of the core clock, while the
execute-stage logic and the PWM generators update on the positive edge.
A hardware-written register captured on one edge is therefore stable and
readable by the pipeline on the other, so a value driven into a register
by hardware becomes visible to an instruction that reads that register
in the same clock period. This half-cycle relationship is the mechanism
by which the register-mapped inputs of Section~\ref{sec:regmap} behave as
ordinary readable registers.

\subsection{Register-Mapped Peripheral Input Interface}
\label{sec:regmap}

\textbf{Problem.} A control loop reads its sensors on every iteration. On
this core a sensor read must ultimately place a value into a register,
because only registers can be arithmetic operands.

\textbf{Conventional approach.} The standard solution is memory-mapped
I/O: the peripheral is assigned an address, and software issues a load to
that address to bring the current value into a register before using it.

\textbf{Drawbacks.} A memory-mapped read is always a distinct load
instruction. On this pipeline that load also exposes the load-use
interlock (Section~\ref{sec:baseline-hazards}): if the loaded value is
used by the next instruction, the read costs a stall cycle in addition to
the load itself. The value must be resident in a register before it can
participate in any computation; there is no way to make the peripheral
value an operand directly.

\textbf{Chosen design.} We map two peripheral inputs directly into
architectural register state. Registers $\$1$ and $\$2$ are implemented
as real 32-bit flip-flop banks---not combinational muxes of a live
input---each with a dedicated write-data input and a write-enable tied
permanently high. Because the enable is always asserted, the two banks
re-capture their peripheral inputs on every clock edge; each register
holds a stable, one-cycle-registered snapshot of its input rather than a
transparent pass-through of a live wire. Reads use the ordinary register
read path: the read decoder selects the stored flop output for $\$1$ or
$\$2$ exactly as it does for any other register. From the software side,
$\$1$ and $\$2$ are simply registers whose contents happen to track the
peripherals.

\begin{figure}[t]
  \centering
  \resizebox{\columnwidth}{!}{
\begin{tikzpicture}[scale=0.64, transform shape,
                    font=\footnotesize, node distance=1.2mm]

  \def\rw{12mm}   
  \def\rh{5.2mm}  
  \tikzset{
    cell/.style={draw=black!55, fill=black!4, minimum width=\rw,
                 minimum height=\rh, inner sep=1pt, font=\footnotesize},
    hcell/.style={cell, draw=accenthw, fill=accenthw!16,
                  postaction={pattern=north east lines, pattern color=accenthw!45}},
    scell/.style={cell, draw=accentsw, fill=accentsw!16},
    ell/.style={font=\footnotesize, minimum height=3.2mm, inner sep=1pt},
    stub/.style={draw=accentsw, thick},
    ocirc/.style={circle, draw=accentsw, thick, fill=white,
                  inner sep=0pt, minimum size=2.1mm}}

  \node[cell]  (g0)  {\$0};
  \node[hcell, below=of g0]  (g1)  {\$1};
  \node[hcell, below=of g1]  (g2)  {\$2};
  \node[cell,  below=of g2]  (g3)  {\$3};
  \node[ell,   below=of g3]  (e1)  {$\vdots$};
  \node[cell,  below=of e1]  (g25) {\$25};
  \node[scell, below=of g25] (g26) {\$26};
  \node[scell, below=of g26] (g27) {\$27};
  \node[scell, below=of g27] (g28) {\$28};
  \node[scell, below=of g28] (g29) {\$29};
  \node[ell,   below=of g29] (e2)  {$\vdots$};
  \node[cell,  below=of e2]  (g31) {\$31};

  \node[bigbox, fill=none, fit=(g0)(g31), inner sep=4pt] (rf) {};
  \node[lbl, rotate=90, anchor=south] at (rf.west) {Register File (32 entries)};

  \coordinate (decW) at ($(rf.east)+(46mm,0)$);   
  \node[block, fill=black!5, minimum width=20mm, minimum height=11mm,
        anchor=west] (dec) at (decW |- g2) {5$\to$32\\decoder};
  \node[block, fill=black!5, minimum width=20mm, above=8mm of dec] (wb)
        {Writeback\\stage};
  \draw[bus] (wb) -- node[note, right, text width=16mm, xshift=1mm]
        {write data\\+\,5-bit index} (dec);

  \draw[sig] (dec.south) |- (g25.east);
  \node[note, text width=24mm, anchor=south west]
        at ($(g25.east)+(6mm,1.5mm)$) {enable[0], enable[3..31]};

  \coordinate (openx) at ($(rf.east)+(10mm,0)$);
  \draw[stub] (dec.west |- g1) -- (openx |- g1) coordinate (stub1end);
  \node[ocirc] at (stub1end) {};
  \draw[stub] (dec.west |- g2) -- (openx |- g2) coordinate (stub2end);
  \node[ocirc] at (stub2end) {};
  \node[note, text=accentsw, text width=30mm, anchor=south west]
        at ($(stub1end)+(3mm,2mm)$)
        {enable[1], enable[2] \emph{unconnected}\\--\,software writes cannot\\reach \$1/\$2};

  \node[block, fill=accenthw!8, left=16mm of g1, minimum width=20mm]
        (dj) {data\_joystick};
  \draw[fwd] (dj) -- (g1);
  \node[note, text=accenthw, anchor=south] at ($(dj.east)!0.5!(g1.west)+(0,1mm)$)
        {enable\,=\,1'b1};

  \node[block, fill=accenthw!8, left=16mm of g2, minimum width=20mm]
        (dg) {data\_gyroscope};
  \draw[fwd] (dg) -- (g2);
  \node[note, text=accenthw, anchor=north] at ($(dg.east)!0.5!(g2.west)+(0,-1mm)$)
        {enable\,=\,1'b1};

  \foreach \r/\i in {g26/0, g27/1, g28/2, g29/3} {
    \draw[bus, draw=accentsw] (\r.east) -- ++(22mm,0)
          node[block, right, fill=accentsw!8, minimum width=11mm] (pwm\i) {PWM\,\i};
  }
  \node[note, text=accentsw, text width=30mm, anchor=north]
        at ($(pwm3.south west)!0.5!(pwm3.south east)+(0,-2mm)$)
        {continuous read exports\\(always driven, not muxed)};

  \node[legend, text width=66mm, anchor=north, font=\scriptsize\itshape]
        at ($(rf.south |- pwm3.south)+(6mm,-12mm)$)
        {hardware ownership enforced structurally:
         no conflict is representable};

\end{tikzpicture}}
  \caption{Register file with three independent write domains. The
  conventional writeback port drives $\$0$ and $\$3$--$\$31$ through a
  5-to-32 decoder; the decoder's enable outputs for indices~1 and~2 are
  left unconnected (open stubs), so software cannot reach the
  hardware-owned registers. Dedicated always-enabled ports drive $\$1$
  and $\$2$ from their peripheral inputs, and $\$26$--$\$29$ are exported
  continuously to the PWM units. No write conflict is representable.}
  \label{fig:regfile}
\end{figure}
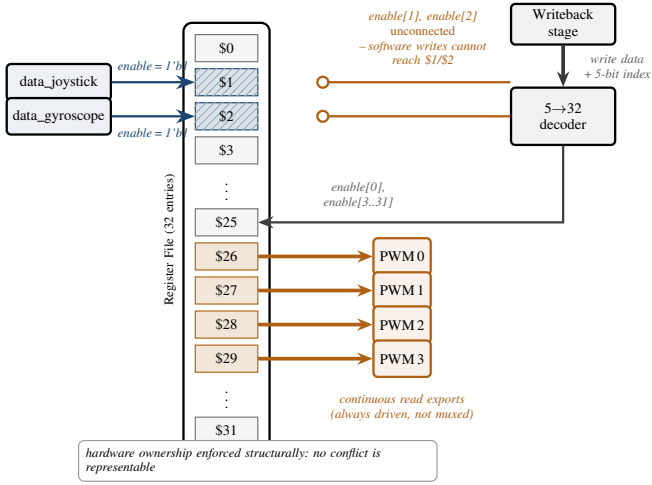

This creates three independent write domains in the register file, shown
in Figure~\ref{fig:regfile}: the conventional single write port, which
drives register~$\$0$ and $\$3$ through $\$31$ via a 5-to-32 decoder; a
dedicated port that drives $\$1$ from its peripheral input; and a
dedicated port that drives $\$2$ from its peripheral input. Hardware
ownership of $\$1$ and $\$2$ is enforced structurally rather than by
arbitration. The conventional write decoder still \emph{generates}
enable outputs for indices~1 and~2, but those enable nets are left
unconnected: no wire carries them into the $\$1$/$\$2$ banks, which
listen only to their tied-high hardware enables. A software writeback
targeting $\$1$ or $\$2$ is therefore not merely ignored at run time; it
is structurally unrepresentable, because no path exists from the
software write port to those two registers. There is no arbiter and no
priority logic because no conflict can occur.

\textbf{Consequences.} A sensor value is now an ordinary register
operand. Reading it requires no load instruction and no interlock cycle;
the value can be named directly in the operand field of whatever
arithmetic instruction consumes it. Section~\ref{sec:evaluation}
quantifies this: in the control loop, peripheral reads cost zero
dedicated instructions and zero cycles. The structural ownership makes
the design correct by construction, at the cost of
permanently dedicating two registers---a cost that, as
Section~\ref{sec:baseline-isa} argued, is zero on a core with no ABI.

\subsection{Hardware PWM Actuation}
\label{sec:pwm}

\textbf{Problem.} The platform's actuators---servos and a motor
driver---are commanded by pulse-width-modulated signals with a fixed
20~ms frame. The control loop must maintain four such waveforms
continuously while it computes.

\textbf{Conventional approach.} On a processor with a timer and an
interrupt controller, periodic waveforms are generated in software: a
timer interrupt fires at the required resolution and a service routine
toggles output pins to build each pulse. This core has neither. Its only
exception sources are arithmetic---overflow and division by zero---so
there is no timer tick to hang a service routine on. The software
baseline available here is therefore not interrupt-driven but
\emph{polled}: the control loop itself must carry four deadline counters
and toggle the output pins inline, interleaving waveform maintenance
with control computation on every iteration.

\textbf{Drawbacks.} Polled waveform generation couples actuation timing
to instruction timing, and does so more tightly than an interrupt-driven
implementation would. Because edges can only be emitted where the loop
happens to test its counters, edge placement is quantized to the loop
period rather than to the timer resolution: a pulse edge lands, in the
worst case, one whole loop iteration late. That quantization appears
directly as pulse-width jitter. The bookkeeping also consumes cycles on
every iteration and complicates the control code with timing logic that
has nothing to do with control.

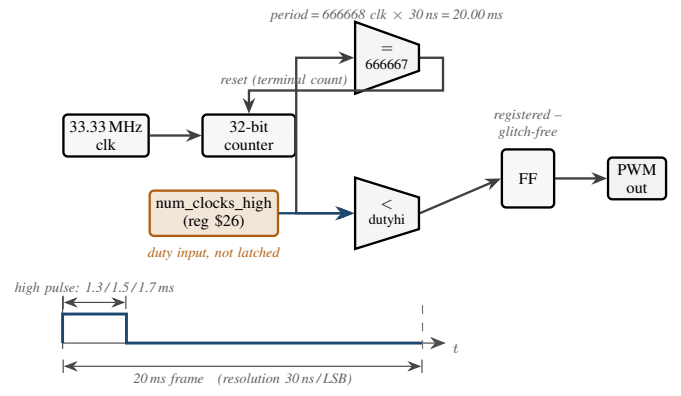
\begin{figure}[t]
  \centering
  \resizebox{\columnwidth}{!}{
\begin{tikzpicture}[scale=0.70, transform shape,
                    font=\footnotesize, node distance=6mm and 9mm]

  \tikzset{
    cmp/.style={draw, thick, fill=black!4, trapezium,
                trapezium angle=64, trapezium stretches body,
                shape border rotate=270, minimum width=9mm,
                minimum height=11mm, inner sep=1pt, font=\scriptsize,
                align=center},
    ff/.style={block, minimum width=9mm, minimum height=10mm}}

  \node[block, align=center] (clk) {33.33\,MHz\\clk};
  \node[block, right=of clk, minimum width=16mm] (cnt) {32-bit\\counter};
  \draw[sig] (clk) -- (cnt);

  \node[cmp, above right=4mm and 10mm of cnt] (cmpP) {$=$\\666667};
  \draw[sig] (cnt.east) |- (cmpP.west);
  \draw[sig] (cmpP.east) -- ++(4mm,0) |- ($(cnt.north)+(0,4mm)$) -- (cnt.north)
        node[note, above, pos=0.15, xshift=6mm]{reset (terminal count)};
  \node[note, above=0.5mm of cmpP, text width=42mm]
        {period\,=\,666668 clk $\times$ 30\,ns\,=\,20.00\,ms};

  \node[cmp, below right=4mm and 10mm of cnt] (cmpD) {$<$\\dutyhi};
  \draw[sig] (cnt.east) |- (cmpD.west);
  \node[block, left=13mm of cmpD, fill=accentsw!16, draw=accentsw,
        minimum width=20mm, align=center] (nch)
        {num\_clocks\_high\\(reg \$26)};
  \draw[fwd] (nch.east) -- (cmpD.west |- nch.east);
  \node[note, text=accentsw, below=0.5mm of nch, text width=30mm]
        {duty input, not latched};

  \node[ff, right=14mm of cmpD, yshift=6mm] (ff) {FF};
  \draw[sig] (cmpD.east) -- (ff.west);
  \node[block, right=of ff, fill=black!5] (out) {PWM\\out};
  \draw[sig] (ff) -- (out);
  \node[note, above=0.5mm of ff, text width=26mm]
        {registered --\\glitch-free};

  \begin{scope}[shift={($(clk.south west)+(0,-32mm)$)}, y=1mm]
    \def\W{62mm}   
    \def\hi{5mm}   
    \draw[->, thin, draw=black!70] (0,0) -- (\W+4mm,0)
          node[note, right, yshift=-1mm]{$t$};
    \draw[thin, draw=black!70] (0,-1mm) -- (0,\hi+3mm);

    \def\pw{11mm}  
    \draw[line width=1pt, draw=accenthw]
      (0,0) -- (0,\hi) -- (\pw,\hi) -- (\pw,0) -- (\W,0);
    \draw[thin, dashed, draw=black!55] (\W,-1mm) -- (\W,\hi+3mm);

    \draw[dimline] (0,\hi+2mm) -- (\pw,\hi+2mm);
    \node[note, anchor=south] at (\pw*0.5,\hi+2mm)
      {high pulse: 1.3\,/\,1.5\,/\,1.7\,ms};
    \draw[dimline] (0,-4mm) -- (\W,-4mm);
    \node[note, anchor=north] at (\W*0.5,-4mm)
      {20\,ms frame \; (resolution 30\,ns\,/\,LSB)};
  \end{scope}

\end{tikzpicture}}
  \caption{One hardware PWM generator. A 32-bit counter clocked at
  33.33~MHz is reset at terminal count 666667; the 666668-clock sequence
  at 30~ns per clock is a 20.00~ms frame. A duty comparator drives the
  registered (glitch-free) output high while the counter is below the
  duty value read live from the channel register ($\$26$ shown); duty
  resolution is 30~ns per LSB. The inset shows one frame with the
  1.3/1.5/1.7~ms servo pulse widths.}
  \label{fig:pwm}
\end{figure}

\textbf{Chosen design.} We offload actuation to four hardware PWM
generators, one per channel, driven by registers $\$26$ through $\$29$.
Registers $\$26$--$\$29$ are ordinary writable registers: software
writes them through the normal write port. They are also
exported as continuous, always-driven read outputs to the PWM units
(they are wired out directly, not selected through a read mux). Each
generator, shown in Figure~\ref{fig:pwm}, is a 32-bit counter clocked at
33.33~MHz. A period comparator resets the counter at a terminal count of
666667, so the counter sequence 0 through 666667 spans 666668 clocks;
at 30~ns per clock this is a 20.00~ms frame. A duty comparator drives
the output high while the counter is below the value in the channel's
register, giving a duty resolution of 30~ns per least-significant bit.
The output is a registered flip-flop, so it is glitch-free. The four
generators drive four board pins (Pmod header~JD); the design constraint
file constrains only the 100~MHz input clock.

\textbf{Consequences.} Actuation leaves the software timeline entirely.
There are no deadline counters in the loop and no software waveform
maintenance; commanding a servo is a single register write, and the
hardware maintains the waveform indefinitely thereafter. The servo pulse
widths used by the platform---1.3, 1.5, and 1.7~ms---are simply the
corresponding counts written to the duty registers. One caveat follows
from the design as built: the duty input is not latched inside the
generator, but read live from the register. Rewriting a duty register in
the middle of a frame changes the comparison threshold immediately, so
the in-flight pulse is perturbed rather than updated cleanly at the next
frame boundary. In the control loop this is benign because duty values
change slowly relative to the frame, but it is a real property of the
design and we return to it in Section~\ref{sec:discussion}.

\subsection{Sensor Input Interface}
\label{sec:sensor}

The two hardware-written registers carry the loop's inputs. Register
$\$1$ carries a 4-bit command input, zero-extended into the 32-bit
register. Register $\$2$ carries a 16-bit sensor word, also zero-extended,
so its upper half is always zero.

In the archived build analyzed here, these inputs come from the
development board: $\$2$ is driven by the 16 board slide switches and
$\$1$ by 4 push buttons, a bring-up configuration. The deployed system
instead drove $\$2$ from an off-chip sensor acquisition path---an inertial
measurement unit read over I2C---whose result reached the processor as a
16-bit parallel word and was captured into $\$2$ by the dedicated hardware
write port. Neither build contains an on-chip I2C master; in both, the
register-mapped interface sees a 16-bit parallel word.

\section{Evaluation}
\label{sec:evaluation}

We evaluate the specializations on the reaction-wheel controller's inner
loop---a proportional/derivative regulator~\cite{astrom2008feedback} that reads
the sensor, decodes the operator command, computes an actuation value,
and writes the PWM registers. This section establishes three results: the peripheral
interface is free at the instruction level; a memory-mapped equivalent
costs five instructions and five cycles per iteration ($11.6\%$ of the
integrated loop, about $5.5\%$ of the archived one); and the loop clears
its 20~ms deadline by roughly $7{,}300\times$ in the archived
configuration and $15{,}000\times$ in the integrated one.

\subsection{Methodology}
\label{sec:eval-method}

We analyze two hardware configurations that differ only in the
multiplier, and use these labels throughout.

The \emph{archived configuration} is a single archived source tree that
contains the peripheral extensions of Section~\ref{sec:extensions}, the
off-chip sensor path of Section~\ref{sec:sensor}, and the core's original
multiplier: a sequential radix-4 Booth unit that stalls the pipeline for
about 16 cycles per multiply (Section~\ref{sec:baseline-alu}). Its
numbers can be recomputed from that one tree, so we report it as the
primary result.

The \emph{integrated configuration} replaces the Booth unit with the
single-cycle array multiplier and is otherwise identical. It matches the
deployed system, but no single archived build contains the extensions and
the single-cycle multiplier together, so we analyze it rather than time
it.

Because these elements were developed in parallel, the numbers below are
a static cycle analysis of the control program, not measurements of one
timed bitstream. The cycle model uses the pipeline behavior established
in Section~\ref{sec:baseline}: one instruction per cycle in steady state,
a two-cycle squash on every taken control transfer, and a one-cycle
load-use interlock. The two configurations share this model and differ
only in multiply latency: about 16 cycles per multiply in the archived
configuration, one cycle in the integrated one. We report cycle counts to
enable a reader to recompute them; we do not report measured wall-clock
timing.

We also did \emph{not} measure resource utilization. We report no LUT,
flip-flop, or DSP counts, and we do not estimate them; the cost of the
specialization in fabric area is outside what the available artifacts let
us state.

\subsection{The Control Loop and Peripheral Accesses}
\label{sec:eval-loop}

The loop reads $\$1$ and $\$2$ a total of ten times (static reads) and
writes three of the four PWM channels---$\$26$, $\$27$, and $\$28$; the
fourth channel, $\$29$, is left at its setup value. The controller's
gains, as coded, are $K_p = 1$ and $K_d = 0$; there is no integral term.
With $K_d = 0$ the derivative term is inert as written: the control law
reduces to pure proportional action, and the multiply that computes the
derivative term is a multiply by zero. One of the three counted
multiplies is therefore dead code, so the loop as measured includes work
a tuned controller would not perform. We report the loop as coded rather
than as intended.

The central result concerns \emph{where the peripheral accesses appear in
the cycle budget}, and the answer is that they do not appear as a
separate cost at all. Because $\$1$ and $\$2$ are ordinary registers,
each of the ten reads is an operand field inside an instruction that
executes anyway---an \texttt{and} that masks a button bit, a
\texttt{sub} that forms the control error, an \texttt{sra} that scales a
term. There is no distinct read instruction to count. One of the PWM
writes is likewise fused: the value written to $\$28$ is the destination
of the \texttt{mul} that computes it, so that write, too, costs no
dedicated instruction. At the instruction level the register-mapped
interface is free, and a memory-mapped interface
structurally cannot match this, because it requires a distinct load to
move a peripheral value into a register before the value can be used.

\begin{table}[t]
  \centering
  \caption{Worst-case control-loop iteration by block, and the cost added
  by a memory-mapped (MMIO) equivalent. The cycle columns depict the
  integrated configuration (single-cycle multiplier), where
  cycles = instruction-cycles + taken-branch penalty, and its MMIO
  counterfactual, which grows only Blocks~1 and~4, to 48 cycles
  ($+11.6\%$). The archived configuration (sequential Booth multiplier)
  differs only in Block~3: its three multiplies each stall the pipeline
  about 16 cycles, adding 48 cycles, so Block~3 is 63 rather than 15 and
  the loop total is 91.}
  \label{tab:blocks}
  \small
  \setlength{\tabcolsep}{4pt}
  \begin{tabular}{@{}lccc@{}}
    \toprule
    Block & Instr & Pen. & Cycles \\
          &       &      & (int/MMIO) \\
    \midrule
    1. Peripheral reads (10$\times$ $\$1$,$\$2$) & 0 & 0 & 0 / 2 \\
    2. Command decode + branch      & 10 & 8 & 18 / 18 \\
    3. P/D compute (3 muls)         & 15 & 0 & 15 / 15 \\
    4. PWM writes ($\$26$, $\$27$)  & 2 & 0 & 2 / 5 \\
    5. Loop-back overhead           & 4 & 4 & 8 / 8 \\
    \midrule
    \textbf{Total}                  & \textbf{31} & \textbf{12} & \textbf{43 / 48} \\
    \bottomrule
  \end{tabular}
\end{table}

Table~\ref{tab:blocks} breaks the integrated configuration's 43-cycle
worst-case iteration into five blocks. The 43 cycles are 31
instruction-cycles plus 12 taken-branch penalty cycles. Block~1---all ten
peripheral reads---contributes zero instructions and zero cycles, exactly
as the argument above predicts, and it does so in both configurations:
the choice of multiplier cannot change whether reading a peripheral needs
its own instruction. The zero-instruction sensor read is therefore
independent of the hardware configuration.

Two observations follow from the table. First, all 12 penalty cycles are
taken-branch squashes: 8 in the command-decode block and 4 in the
loop-back. Branch overhead ($12$ cycles, about $28\%$ of the integrated
loop) thus dwarfs anything the peripheral interface costs, which
reinforces the reading that the specialization optimized a part of the
loop that was never the bottleneck. This 12-cycle figure assumes an
unconditional jump squashes two cycles exactly as a conditional branch
does; three of the loop's jumps are redundant fall-through no-ops that a
peephole optimizer would remove, either of which would lower it. Second,
the multiplier sets the loop length. In the archived configuration the
three multiplies in Block~3 use the sequential Booth unit, which stalls
the pipeline about 16 cycles each; the three together add 48 cycles, so
Block~3 costs 63 cycles and the loop runs 91. Substituting the
single-cycle array multiplier reduces Block~3 from 63 to 15 cycles and
the loop from 91 to 43---the integrated configuration. The multiplier,
not the peripheral interface, sets the loop length in both cases.

\subsection{Memory-Mapped Counterfactual}
\label{sec:eval-mmio}

To price the register mapping, we ask what the same loop would cost under
MMIO on the same core. A competent implementation keeps a peripheral base
pointer in a register (set up once, amortized to zero per iteration),
snapshots each input once per iteration rather than reloading it at every
use, and schedules the loads so their results are not needed in the very
next instruction, hiding the load-use interlock. Under those best-case
assumptions, only two blocks of Table~\ref{tab:blocks} grow. Block~1
gains two loads to snapshot $\$1$ and $\$2$ (0 to 2 cycles), and Block~4
gains a store for each PWM channel written this iteration---$\$26$,
$\$27$, and $\$28$---growing from 2 to 5 cycles. The $\$28$ value is
produced by the fused multiply in Block~3, but MMIO cannot fold its store
into that multiply and needs a separate instruction. Blocks~2, 3, and~5
are unchanged. The counterfactual adds 5 instructions and 5 cycles per
iteration: $11.6\%$ of the 43-cycle integrated loop, and about $5.5\%$ of
the 91-cycle archived loop. Figure~\ref{fig:loop_timing_mmio} draws the
integrated configuration against its counterfactual block by block,
isolating where the cost appears.

A naive MMIO version that reloads each peripheral value on every use,
back to back, would cost more ($+13$ instructions and $+23$ cycles), but
that larger number is an artifact of poor scheduling: a competent
programmer avoids it by hoisting the load out of the interlock shadow. We
therefore do not present the larger figure as the cost of MMIO. The cost
of a memory-mapped interface on this core is five instructions and five
cycles per iteration, and it arises entirely because MMIO cannot make a
peripheral value a direct operand.

\begin{figure*}[t]
  \centering
  \resizebox{\textwidth}{!}{
\begin{tikzpicture}[font=\footnotesize]

  \def\Wmmio{130}                 
  \pgfmathsetmacro{\cyc}{\Wmmio/48}   
  \def\barh{8mm}\def\barhalf{4mm}
  \def\gapy{16mm}                 

  \tikzset{
    bseg/.style={draw=black!60, line width=0.5pt, minimum height=\barh,
                 inner sep=0pt},
    growseg/.style={bseg, fill=accentsw!30,
                 postaction={pattern=north east lines,
                             pattern color=accentsw!55}},
    keepA/.style={bseg, fill=accenthw!14},
    keepB/.style={bseg, fill=black!8},
    keepC/.style={bseg, fill=accenthw!30},
    seglbl/.style={font=\scriptsize, align=center},
    growtag/.style={font=\scriptsize\bfseries, text=accentsw, align=center}}

  \newcommand{\drawseg}[4]{%
    \path[#4] (#2*\cyc mm, #1-\barhalf) rectangle (#3*\cyc mm, #1+\barhalf);}

  \def\yR{0mm}
  \drawseg{\yR}{0}{18}{keepA}     
  \drawseg{\yR}{18}{33}{keepB}    
  \drawseg{\yR}{33}{35}{growseg}     
  \drawseg{\yR}{35}{43}{keepC}    
  \draw[accentsw, line width=1.3pt] (0,\yR-\barhalf-1mm) -- (0,\yR+\barhalf+1mm);
  \node[circle, draw=accentsw, fill=accentsw!18, line width=0.7pt,
        inner sep=0.3pt, minimum size=3.8mm, font=\scriptsize\bfseries,
        text=accentsw] at (0,\yR+\barhalf+3.5mm) {0};
  \node[seglbl, anchor=east] at (-4mm,\yR) {\textbf{Reference}\\(register-mapped)};
  \node[seglbl] at (9*\cyc mm,\yR) {joystick\\decode+br};
  \node[seglbl] at (25.5*\cyc mm,\yR) {P/D compute};
  \node[seglbl] at (39*\cyc mm,\yR) {loop-back};

  \def\yM{-16mm}                     
  \drawseg{\yM}{0}{2}{growseg}       
  \drawseg{\yM}{2}{20}{keepA}     
  \drawseg{\yM}{20}{35}{keepB}    
  \drawseg{\yM}{35}{40}{growseg}     
  \drawseg{\yM}{40}{48}{keepC}    
  \node[seglbl, anchor=east] at (-4mm,\yM) {\textbf{MMIO}\\(load/store)};

  \node[note, anchor=north] at (9*\cyc mm, \yR-\barhalf) {18};
  \node[note, anchor=north] at (25.5*\cyc mm, \yR-\barhalf) {15};
  \node[note, anchor=north] at (39*\cyc mm, \yR-\barhalf) {8};
  \node[note, anchor=north] at (11*\cyc mm, \yM-\barhalf) {18};
  \node[note, anchor=north] at (27.5*\cyc mm, \yM-\barhalf) {15};
  \node[note, anchor=north] at (44*\cyc mm, \yM-\barhalf) {8};

  \node[note, anchor=north] at (34*\cyc mm, \yR-\barhalf) {2};   
  \node[note, anchor=north] at (37.5*\cyc mm, \yM-\barhalf) {5}; 
  \node[note, anchor=north] at (1*\cyc mm, \yM-\barhalf) {2};    

  \coordinate (sgrow) at (1*\cyc mm, \yM+\barhalf);   
  \coordinate (pgrow) at (37.5*\cyc mm, \yM+\barhalf); 
  \node[growtag, anchor=south west, text width=26mm] (gsens)
        at (3*\cyc mm, \yR+\barhalf+3mm) {sensor reads: 0\,$\to$\,2\,cyc};
  \draw[fwd] (gsens.south west) |- (sgrow);
  \node[growtag, anchor=south, text width=24mm] (gpwm)
        at (37.5*\cyc mm, \yR+\barhalf+3mm) {PWM writes: 2\,$\to$\,5\,cyc};
  \draw[fwd] (gpwm.south) -- (pgrow);

  \draw[dashed, black!55] (43*\cyc mm, \yR+\barhalf+2mm) -- (43*\cyc mm, \yM-\barhalf-8mm);
  \draw[decorate, decoration={brace, amplitude=5pt, mirror},
        draw=accentsw, thick]
        (43*\cyc mm, \yM-\barhalf-9mm) -- (48*\cyc mm, \yM-\barhalf-9mm);
  \node[growtag, anchor=north, text width=34mm]
        at (45.5*\cyc mm, \yM-\barhalf-13mm)
        {+5 cycles (+11.6\%)};

  \node[note, anchor=west] at (43*\cyc mm, \yR) {= 43 cyc};
  \node[note, anchor=west] at (48*\cyc mm, \yM) {= 48 cyc};

  \node[note, anchor=north west, text width=118mm]
        at (0mm, \yM-\barhalf-20mm)
        {Only the \textbf{\textcolor{accentsw}{sensor-read}} and
         \textbf{\textcolor{accentsw}{PWM-write}} blocks grow: MMIO needs
         explicit load/store instructions where the reference design uses
         register operands.  The other three blocks are unchanged.};

\end{tikzpicture}}
  \caption{Per-block comparison of the integrated configuration
  (single-cycle multiplier) against its memory-mapped counterfactual.
  Only two blocks grow: the sensor reads, from zero to two cycles for a
  pair of loads that snapshot the peripheral values once per iteration,
  and the actuation writes, by three cycles for a store after each PWM
  update ($\$26$, $\$27$, and $\$28$). The decode, computation, and
  loop-back blocks are unchanged. The total difference is five cycles, or
  $11.6\%$ of the 43-cycle integrated iteration (about $5.5\%$ of the
  91-cycle archived iteration).}
  \label{fig:loop_timing_mmio}
\end{figure*}

\subsection{Timing Headroom}
\label{sec:eval-headroom}

\begin{figure*}[t]
  \centering
  \resizebox{\textwidth}{!}{
\begin{tikzpicture}[font=\footnotesize]

  \def\Wtopnum{132}   
  \def\barh{7mm}      
  \def\barhalf{3.5mm} 
  \tikzset{
    seg/.style={draw=black!60, fill=black!6, minimum height=\barh,
                inner sep=1pt, font=\scriptsize, align=center},
    segA/.style={seg, fill=accenthw!14},
    segB/.style={seg, fill=accentsw!16},
    penband/.style={draw=black!55, line width=0.4pt,
                    pattern=north east lines, pattern color=black!58},
    zbox/.style={draw=accenthw, fill=accenthw!7, rounded corners=2pt,
                 inner sep=3pt, font=\scriptsize, align=left,
                 text=black!80}}

  \def\cJoy{18}\def\cPD{15}\def\cPWM{2}\def\cLoop{8}
  \def\pJoy{8}\def\pLoop{4}   
  \pgfmathsetmacro{\uJoy}{\Wtopnum*\cJoy/43}
  \pgfmathsetmacro{\uPD}{\Wtopnum*\cPD/43}
  \pgfmathsetmacro{\uPWM}{\Wtopnum*\cPWM/43}
  \pgfmathsetmacro{\uLoop}{\Wtopnum*\cLoop/43}
  \pgfmathsetmacro{\xJoyEnd}{\uJoy}
  \pgfmathsetmacro{\xPDEnd}{\uJoy+\uPD}
  \pgfmathsetmacro{\xPWMEnd}{\uJoy+\uPD+\uPWM}
  \pgfmathsetmacro{\wpJoy}{\Wtopnum*\pJoy/43}
  \pgfmathsetmacro{\wpLoop}{\Wtopnum*\pLoop/43}
  \pgfmathsetmacro{\xJoyPenL}{\xJoyEnd-\wpJoy}
  \pgfmathsetmacro{\xLoopPenL}{\Wtopnum-\wpLoop}

  \coordinate (t0) at (0,0);
  \node[segB, anchor=west, minimum width=\uJoy mm] (s2) at (t0) {};
  \node[segA, anchor=west, minimum width=\uPD mm] (s3) at (s2.east) {};
  \node[segB, anchor=west, minimum width=\uPWM mm] (s4) at (s3.east) {};
  \node[segA, anchor=west, minimum width=\uLoop mm] (s5) at (s4.east) {};

  \path[penband] (\xJoyPenL mm,-\barhalf) rectangle (\xJoyEnd mm,\barhalf);
  \path[penband] (\xLoopPenL mm,-\barhalf) rectangle (\Wtopnum mm,\barhalf);

  \node[font=\scriptsize, align=center] at (s2) {joystick decode\\+ branch dispatch};
  \node[font=\scriptsize, align=center] at (s3) {P/D\\compute};
  \node[font=\scriptsize, align=center] at (s5) {loop-\\back};

  \node[note, below=0.5mm of s2] {18\,cyc};
  \node[note, below=0.5mm of s3] {15\,cyc};
  \node[note, below=0.5mm of s5] {8\,cyc};
  \node[note, anchor=north] (pwmlbl) at ($(s4.south)+(0,-6mm)$)
        {PWM writes\\(\$26,\$27): 2\,cyc};
  \draw[note, ->, >=Stealth, draw=black!55] (pwmlbl.north) -- (s4.south);

  \node[taglbl, anchor=south] at ($(\Wtopnum*0.5 mm,\barhalf)+(0,15mm)$)
        {One control-loop iteration = \textbf{43 cycles} = 1.29\,\textmu s\ \ (30\,ns/cycle)};
  \node[note, anchor=south] at ($(\Wtopnum*0.5 mm,\barhalf)+(0,10.5mm)$)
        {31 instruction cycles + 12 taken-branch penalty cycles = 43};
  \draw[dimline] ($(s2.north west)+(0,7mm)$) -- ($(s5.north east)+(0,7mm)$);

  \draw[accenthw, line width=1.4pt] (0,-\barhalf-1.5mm) -- (0,\barhalf+1.5mm);
  \node[circle, draw=accenthw, fill=accenthw!16, line width=0.8pt,
        inner sep=0.4pt, minimum size=4.6mm, font=\scriptsize\bfseries,
        text=accenthw] (zbadge) at (0,0) {0};
  \node[zbox, anchor=north west, text width=45mm] (zcall)
        at (3mm,-15mm)
        {\textbf{sensor reads: 0 cycles}\\peripheral values are register
         operands --- no access instruction};
  \draw[fwd] (0,-14.5mm) -- (zbadge.south);

  \node[note, anchor=north east, text width=58mm] (hleg)
        at (\Wtopnum mm,-15mm)
        {\tikz\path[penband] (0,0) rectangle (3.4mm,2.4mm);\ \ =\ taken-branch
         penalty (8\,cyc in decode/dispatch, 4\,cyc in loop-back)};

  \coordinate (b0) at ($(s2.south west)+(0,-52mm)$);
  \def\Wframe{\Wtopnum mm}
  \def\sliver{0.9mm}          
  \def\breakx{34mm}           
  \def\afterbreak{40mm}       

  \fill[accenthw!70] (b0) rectangle ($(b0)+(\sliver,\barh)$);
  \draw[accenthw, line width=0.7pt] (b0) rectangle ($(b0)+(\sliver,\barh)$);
  \coordinate (act) at ($(b0)+(\sliver*0.5,\barh)$);   
  \fill[black!6] ($(b0)+(\sliver,0)$) rectangle ($(b0)+(\breakx,\barh)$);
  \draw[black!60] ($(b0)+(0,0)$) rectangle ($(b0)+(\breakx,\barh)$);
  \fill[black!6] ($(b0)+(\afterbreak,0)$) rectangle ($(b0)+(\Wframe,\barh)$);
  \draw[black!60] ($(b0)+(\afterbreak,0)$) rectangle ($(b0)+(\Wframe,\barh)$);

  \foreach \dx in {0mm, 4mm} {
    \draw[white, line width=2.4pt]
      ($(b0)+(\breakx+\dx,-0.6mm)$) -- ($(b0)+(\breakx-1.2mm+\dx,\barh*0.4)$)
      -- ($(b0)+(\breakx+1.2mm+\dx,\barh*0.6)$) -- ($(b0)+(\breakx+\dx,\barh+0.6mm)$);
    \draw[black!60, line width=0.6pt]
      ($(b0)+(\breakx+\dx,-0.6mm)$) -- ($(b0)+(\breakx-1.2mm+\dx,\barh*0.4)$)
      -- ($(b0)+(\breakx+1.2mm+\dx,\barh*0.6)$) -- ($(b0)+(\breakx+\dx,\barh+0.6mm)$);
  }

  \draw[fwd] ($(act)+(0,11mm)$) -- (act);
  \node[note, text=accenthw, anchor=south west, text width=46mm]
        at ($(act)+(-1mm,11mm)$)
        {1.29\,\textmu s of 20\,ms\\= \textbf{0.0065\%} utilization};

  \node[taglbl, anchor=north] at ($(b0)+(\Wframe*0.5,0)+(0,-4mm)$)
        {20\,ms PWM frame};
  \node[note, anchor=south east] at ($(b0)+(\Wframe,\barh)+(0,1mm)$) {frame end};

\end{tikzpicture}}
  \caption{Control-loop timing budget for the integrated configuration
  (single-cycle multiplier). Top: one worst-case iteration,
  43~cycles ($1.29~\mu$s at 30~ns/cycle), split into the blocks of
  Table~\ref{tab:blocks}. Bottom: the same $1.29~\mu$s against the 20~ms
  PWM frame---about $0.0065\%$ of the deadline, drawn as a sliver with a
  broken axis to convey the scale gap. The deadline is met with a margin
  of roughly $15{,}000\times$; the archived configuration's 91-cycle
  iteration meets it by roughly $7{,}300\times$.}
  \label{fig:loop_timing}
\end{figure*}
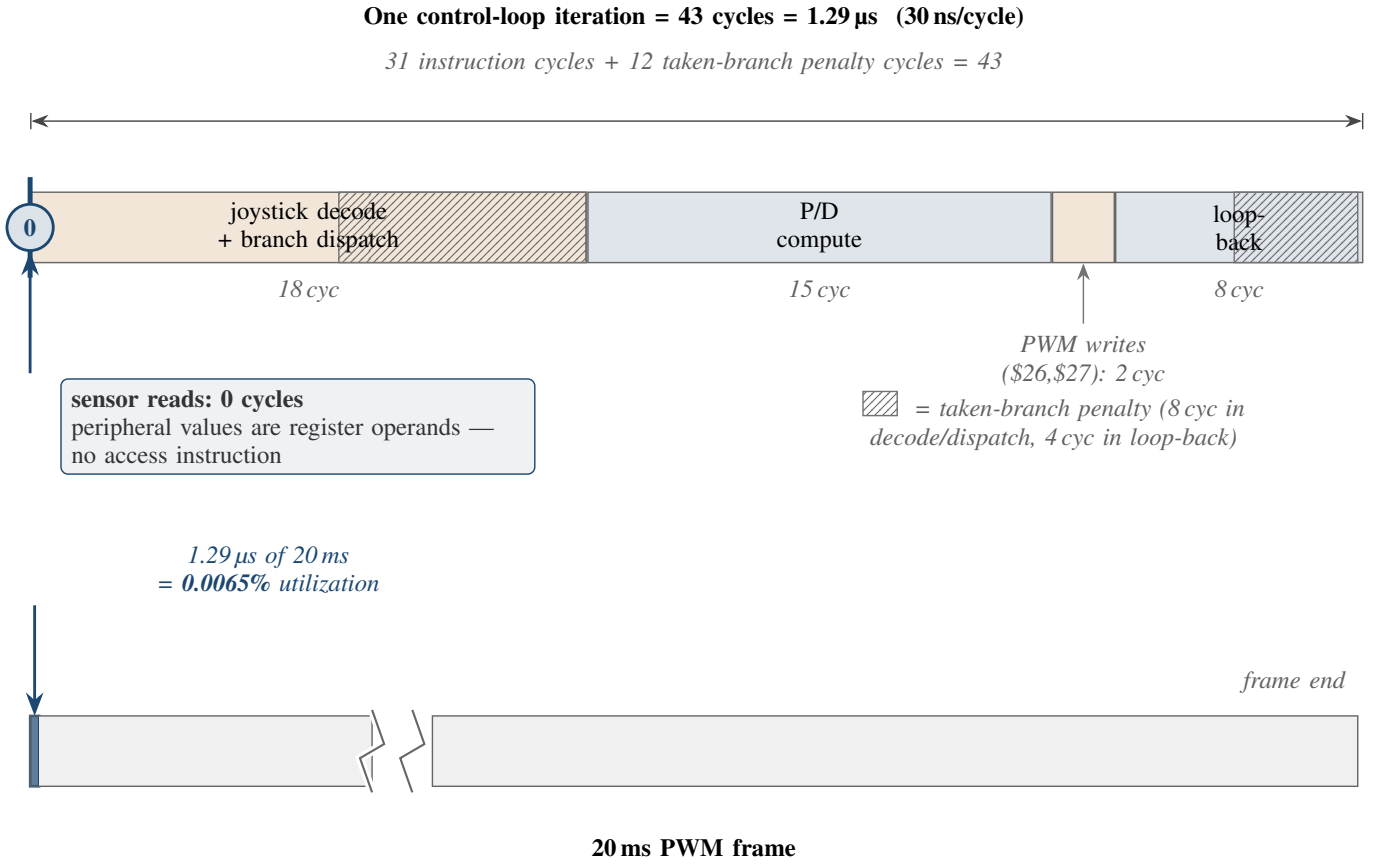

The archived configuration's worst-case iteration is 91 cycles. At the
30~ns clock period that is $91 \times 30~\text{ns} = 2.73~\mu$s, a loop
rate of about 366~kHz. The actuation deadline is one 20~ms PWM frame, so
the loop consumes $2.73~\mu\text{s} / 20~\text{ms} \approx 0.014\%$ of its
deadline---a margin of roughly $7{,}300\times$. The integrated
configuration, with the single-cycle multiplier, completes the same worst
case in 43 cycles: $1.29~\mu$s (about 775~kHz), or $0.0065\%$ of the
deadline, a margin of roughly $15{,}000\times$.
Figure~\ref{fig:loop_timing} draws the integrated iteration against the
frame; the active work is a sliver too thin to see without a broken axis.

The headroom finding does not depend on the configuration. Whether the
loop takes 91 cycles or 43, it clears the 20~ms deadline by three to four
orders of magnitude, so the finding that the specialization was not
necessary for deadline compliance holds either way.

This is the paper's most important finding. The specialization was
not necessary for the loop to meet its deadline. At a margin of three to
four orders of magnitude---in either configuration---an ordinary
memory-mapped implementation on this same core, even the naive one above,
would have met the 20~ms deadline comfortably. The value of the
register mapping is therefore not real-time compliance, and it is not
determinism: an on-chip single-cycle I/O region is already deterministic.
The value is what Sections~\ref{sec:eval-loop} and~\ref{sec:eval-mmio}
measured: one fewer instruction and one fewer potential stall per
access, the resulting reduction in loop size, and the
qualitative software simplification of the actuation offload.

\subsection{Qualitative Software Simplification}
\label{sec:eval-software}

Beyond the cycle counts, the specialization removes whole categories of
software from the loop. Because the PWM hardware maintains the waveforms,
the loop carries no deadline counters and performs no waveform
maintenance. Because peripherals are named as registers
rather than addressed, the loop carries no address arithmetic and no
peripheral base-address register. This is the benefit that does
not reduce to a percentage: the control program contains only control.

\subsection{Bounding the Software Actuation Baseline}
\label{sec:eval-swpwm}

The following is an
analytical bound rather than a measurement. It is worth stating because
it sets the scale of what the actuation offload avoids, and because the
bound is tight enough that measurement would add little.

As Section~\ref{sec:pwm} notes, this core has no timer and no interrupt
controller, so a software implementation must poll: the control loop
carries a deadline counter per channel and toggles the output pins
inline. Two consequences follow directly from the loop period. First,
edge placement is quantized to one loop iteration, because an edge can
only be emitted where the loop tests its counters. A pulse edge is
therefore up to one iteration late---$1.29~\mu$s in the integrated
configuration, $2.73~\mu$s in the archived one---and that quantization
appears as pulse-width jitter. Against the 1.3--1.7~ms pulse widths the
platform uses, this is a jitter of roughly $0.1\%$ to $0.2\%$ of pulse
width. Second, maintaining four channels costs a counter comparison and
a conditional branch per channel per iteration, which on this
branch-expensive core (Section~\ref{sec:eval-loop}) would add cycles
comparable to the command-decode block, plausibly increasing the loop
period by half again.

Neither figure threatens the deadline; a servo will not resolve $2~\mu$s
of jitter in a 20~ms frame, and the loop has three to four orders of
magnitude of headroom to absorb the extra cycles. The hardware PWM's
jitter is zero by construction, so the offload converts a small,
bounded, timing-dependent error into no error at all. That is a real
benefit and a small one, which is the same conclusion the cycle counts
reached by a different route.

\section{Platform}
\label{sec:platform}

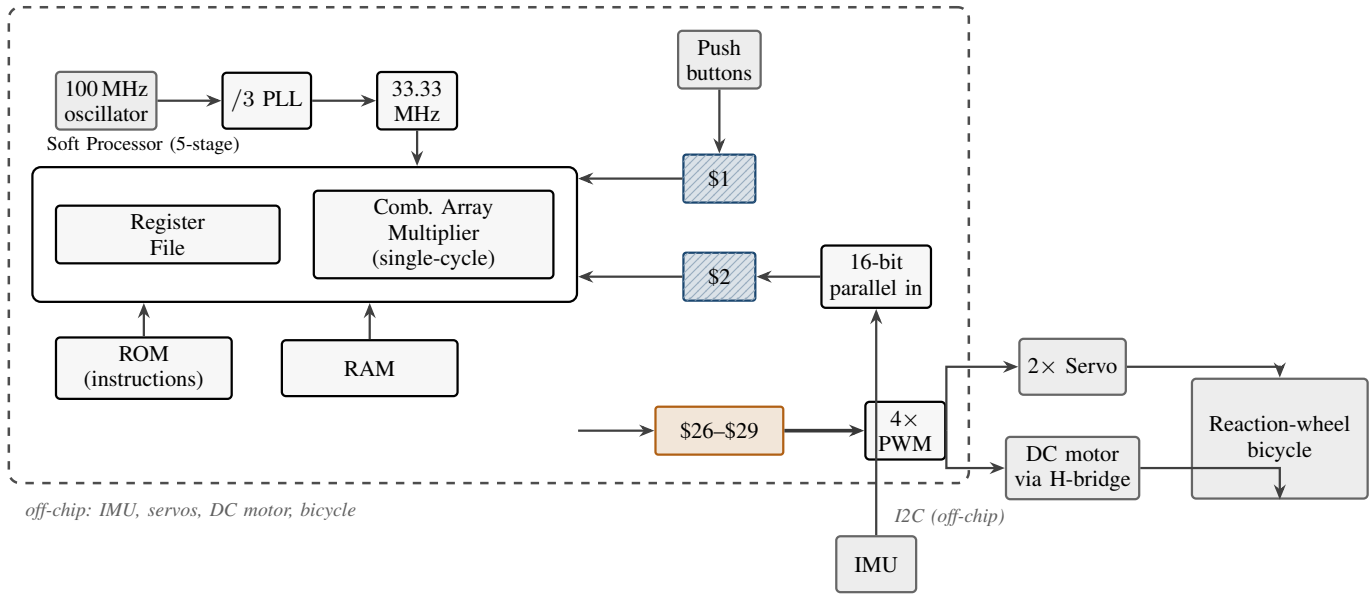
\begin{figure*}[t]
  \centering
  \resizebox{\textwidth}{!}{
\begin{tikzpicture}[font=\footnotesize, node distance=5mm and 8mm]

  \node[periph, align=center] (osc) {100\,MHz\\oscillator};
  \node[block, right=of osc] (pll) {$/3$ PLL};
  \node[block, right=of pll, align=center] (clk) {33.33\\MHz};
  \draw[sig] (osc) -- (pll);
  \draw[sig] (pll) -- (clk);

  \node[block, below=13mm of osc.west, anchor=north west, minimum width=28mm,
        align=center] (rfile) {Register\\File};
  \node[block, right=4mm of rfile, minimum width=30mm, align=center] (mul)
        {Comb. Array\\Multiplier\\(single-cycle)};
  \node[bigbox, fill=none, fit=(rfile)(mul), inner sep=8pt] (cpu) {};
  \node[lbl, anchor=south west, xshift=2pt, yshift=0.5pt] at (cpu.north west)
        {Soft Processor (5-stage)};
  \draw[sig] (clk) -- (clk |- cpu.north);

  \node[block, below=9mm of rfile.south west, anchor=north west,
        minimum width=22mm] (rom) {ROM\\(instructions)};
  \node[block, right=6mm of rom, minimum width=22mm] (ram) {RAM};
  \draw[sig] (rom.north) -- (rom.north |- cpu.south);
  \draw[sig] (ram.north) -- (ram.north |- cpu.south);

  \node[hwreg, right=16mm of mul.east, yshift=7mm] (r1) {\$1};
  \node[hwreg, below=6mm of r1] (r2) {\$2};
  \node[swreg, below=13mm of r2, minimum width=16mm] (rout) {\$26--\$29};
  \draw[sig] (r1.west) -- (r1.west -| cpu.east);
  \draw[sig] (r2.west) -- (r2.west -| cpu.east);
  \draw[sig] (cpu.east |- rout) -- (rout.west);

  \node[periph, align=center, above=8mm of r1] (btn) {Push\\buttons};
  \draw[sig] (btn) -- (r1);
  \node[block, right=8mm of r2, align=center] (i2c) {16-bit\\parallel in};
  \draw[sig] (i2c.west) -- (r2.east);

  \node[block, right=10mm of rout, align=center] (pwm) {4$\times$\\PWM};
  \draw[bus] (rout) -- (pwm);
  \node[periph, right=9mm of pwm, align=center, yshift=8mm] (servo)
        {2$\times$ Servo};
  \node[periph, below=5mm of servo, align=center] (hb)
        {DC motor\\via H-bridge};
  \draw[sig] (pwm.east) |- (servo.west);
  \draw[sig] (pwm.east) |- (hb.west);
  \node[periph, right=8mm of servo, align=center, minimum width=22mm,
        minimum height=15mm, yshift=-9mm] (bike) {Reaction-wheel\\bicycle};
  \draw[sig] (servo.east) -| (bike.north);
  \draw[sig] (hb.east) -| (bike.south);

  \begin{scope}[on background layer]
    \node[fpgabound, fit=(osc)(pll)(clk)(cpu)(rom)(ram)(btn)(i2c)(r1)(r2)(rout)(pwm)]
          (fpga) {};
  \end{scope}
  \node[taglbl, anchor=south west] at ($(fpga.north west)+(0,0.8mm)$)
        {Artix-7 XC7A100T (Nexys A7-100T)};

  \node[periph, align=center] (imu) at ($(i2c |- fpga.south)+(0,-10mm)$) {IMU};
  \draw[sig] (imu) -- node[note, right, pos=0.35, xshift=1mm]{I2C (off-chip)}
        (imu |- fpga.south) -- (i2c.south);
  \node[note, anchor=north west] at ($(fpga.south west)+(1mm,-1mm)$)
        {off-chip: IMU, servos, DC motor, bicycle};

\end{tikzpicture}}
  \caption{Complete system. A 100~MHz oscillator is divided by three to
  the 33.33~MHz core clock; the soft processor holds the register file
  and the single-cycle array multiplier alongside instruction and data
  memory. Push buttons feed $\$1$; the off-chip IMU is acquired over I2C
  and presented as a 16-bit parallel word feeding $\$2$; $\$26$--$\$29$
  drive four hardware PWM channels to two servos and a DC motor. The
  dashed boundary is the Artix-7 XC7A100T; the IMU, servos, motor, and
  bicycle are off-chip.}
  \label{fig:system}
\end{figure*}

The processor drives a reaction-wheel self-balancing
bicycle~\cite{meijaard2007bicycle}, shown as a whole system in
Figure~\ref{fig:system}. The chassis is 3D-printed
(designed in Onshape over seven revisions). A continuous-rotation servo
provides drive and a micro-servo provides steering; a 6:1 geared 12~V
brushed DC motor spins a reaction wheel, a disc flywheel carrying two
bolt rings whose placement adjusts its inertia. Power is a 12~V supply
with a 12-to-5~V regulator and a 3.3-to-5~V level shifter between the
FPGA I/O and the actuators. The reaction-wheel motor was sized from a
torque budget: with a chassis inertia of $51.168~\text{in}^2\,\text{oz}$
and a wheel inertia of $7.536~\text{in}^2\,\text{oz}$, the required wheel
torque is $80.37~\text{in}\,\text{oz}$ at about 101.8~RPM. In the deployed
system the inertial measurement unit is read over I2C; as
Section~\ref{sec:sensor} notes, that acquisition is off-chip and the value
reaches the processor as a 16-bit parallel word mapped to $\$2$. The
archived build analyzed here uses a bring-up configuration instead, in
which $\$2$ is driven by the board slide switches and $\$1$ by push
buttons.

\begin{figure}[t]
  \centering
  \includegraphics[width=\columnwidth]{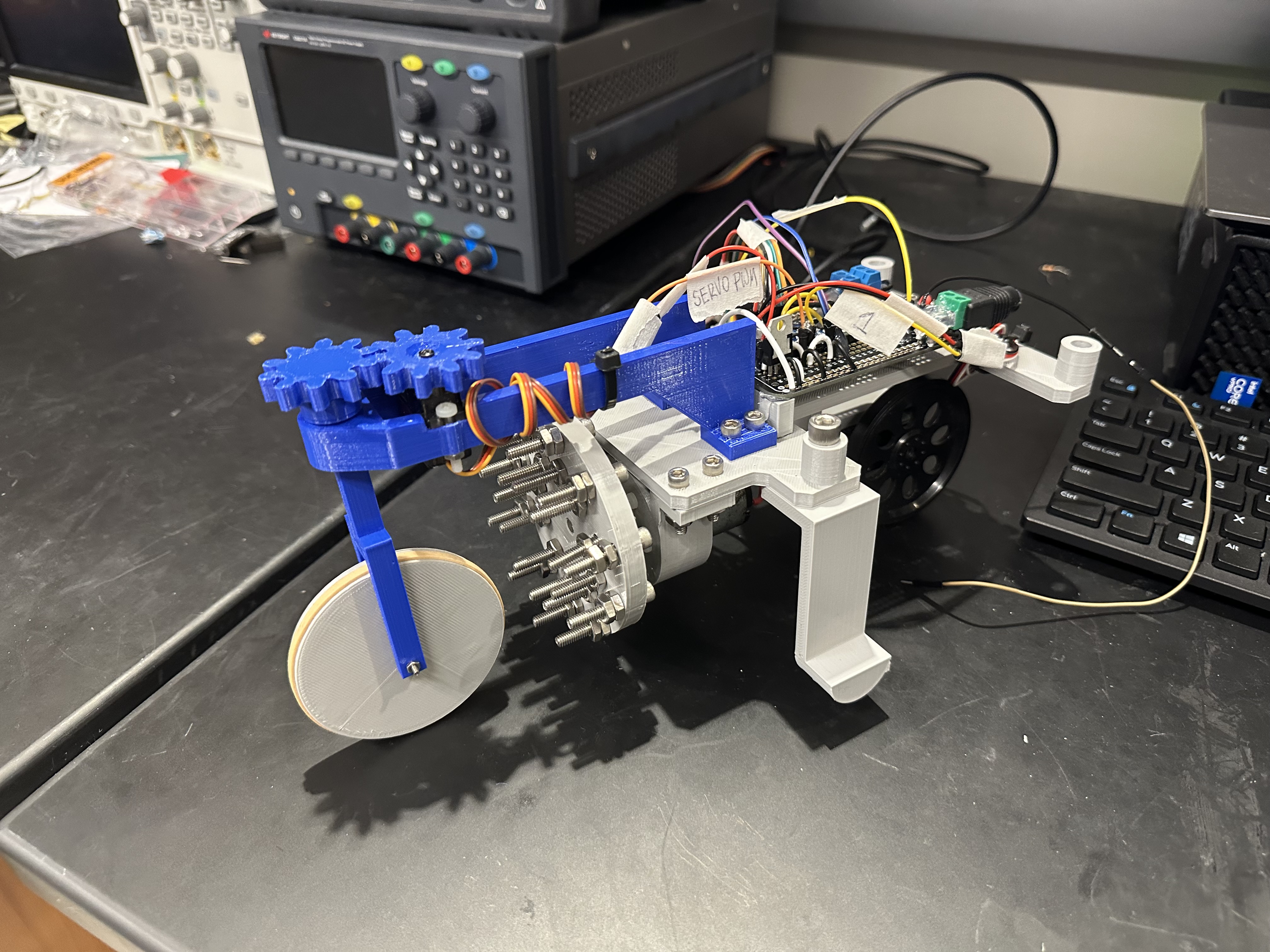}
  \caption{The assembled reaction-wheel bicycle.}
  \label{fig:photo}
\end{figure}

We are explicit about the platform's status: sustained autonomous
balancing was not achieved within the project timeline. The platform
validated the processor's peripheral interface---sensor values arriving
in registers, PWM commands leaving through registers, and the control
loop closing over both---but it did not validate the balancing
controller itself.

\section{Discussion and Limitations}
\label{sec:discussion}

The specialization adds no dedicated instructions to the sensor-read path
and simplifies software, but the design as built has real limitations, and
we set them out plainly.

\textbf{No clock-domain-crossing synchronizer on the input path.} The
peripheral inputs feed the $\$1$/$\$2$ flip-flop banks through a single
capture flop, not a two-flop synchronizer. For inputs that are
synchronous to the core clock this is adequate, but for a genuinely
asynchronous sensor source a single capture flop provides no
metastability protection, and a two-flop synchronizer would be required.
We note that the design does contain a two-flop synchronizer
elsewhere---on the data-memory read path---so the technique was available
and simply not applied to the register-mapped inputs. We list this
synchronizer as future work.

\textbf{No snapshot, valid bit, or double buffer.} Nothing in the
register-mapped path guarantees that a multi-field sensor sample is
coherent: there is no valid bit to mark a complete sample and no double
buffer to hold one sample stable while another arrives. The demonstrated
system sidesteps this rather than solving it, by reading a single axis so
that no multi-field coherency question arises. A design that assembled a
multi-word sample would need explicit snapshot logic that this one does
not have.

\textbf{Unlatched PWM duty.} As Section~\ref{sec:pwm} described, each
generator reads its duty value live rather than latching it at a frame
boundary. A mid-frame rewrite perturbs the in-flight pulse. This is
benign for slowly varying duties but is a correctness hazard for any
control law that updates duties aggressively within a frame.

\textbf{Scalability.} The approach consumes one architectural register
per mapped peripheral, and the ISA has 32 general-purpose registers.
Register mapping therefore caps out at a handful of peripherals, whereas
MMIO scales to thousands of devices behind a single address space. The
technique is thus a targeted optimization for a small number of
frequently accessed peripherals, not a general I/O strategy.

\textbf{Toolchain cost.} A hardware-owned register breaks an invariant
that compilers rely on---that a general-purpose register is stable across
reads and holds what software last wrote. A conventional register
allocator cannot safely allocate $\$1$ or $\$2$, because their contents
change underneath it. This design is viable only because its software is
hand-written assembly with author-defined conventions
(Section~\ref{sec:baseline-isa}); on a core with a compiler and an ABI,
mapping peripherals into allocatable registers would be considerably more
disruptive.

\textbf{Reconstruction and unmeasured cost.} As
Section~\ref{sec:eval-method} stated, the archived and integrated
configurations are analyzed statically rather than measured from a single
archived bitstream, and we did not measure resource utilization. The
cycle-level claims are therefore analytic. The fabric-area cost is
likewise unmeasured and is plausibly non-trivial: the four PWM read ports
are exported as always-on wires and the register file drives an internal
tri-state read bus, adding routing and fan-out that we did not quantify.

\section{Conclusion and Future Work}
\label{sec:conclusion}

We specialized a five-stage soft processor for one real-time control
application by mapping two frequently accessed peripheral inputs into
architectural registers and offloading four periodic actuation channels
to hardware PWM generators. The
register mapping is free at the instruction level: because a sensor value
is an ordinary register operand, the control loop's ten sensor reads
cost no dedicated instruction and no dedicated cycle. A memory-mapped
interface cannot achieve this, because it must load a value before using it.
The memory-mapped equivalent costs five instructions and five cycles per
iteration, and the
actuation offload removes waveform maintenance from software
entirely. Yet the worst-case loop finishes in $2.73~\mu$s in
the archived configuration and $1.29~\mu$s in the integrated one,
against a 20~ms deadline---margins of roughly $7{,}300\times$ and
$15{,}000\times$---so the specialization was not necessary for
real-time compliance. Its value is instruction count and software
simplicity, not determinism, which the on-chip single-cycle I/O region
already provided. This paper's contribution is a
precise account of what the design point buys.

Several extensions follow directly from Section~\ref{sec:discussion}. A
two-flop synchronizer on the $\$1$/$\$2$ input path would make the
interface safe for asynchronous sensor sources. Snapshot registers with a
valid bit would let the interface carry a coherent multi-axis sample
rather than a single axis. Measuring resource utilization would close the
one cost dimension this study left unquantified.

\bibliographystyle{IEEEtran}
\bibliography{references}

\end{document}